# Gibbs Paradox and Similarity Principle

Shu-Kun Lin

*Molecular Diversity Preservation International (MDPI),*
*Matthaeusstrasse 11, CH-4057 Basel, Switzerland; E-mail: lin@mdpi.org*

**Abstract.** As no heat effect and mechanical work are observed, we have a simple experimental resolution of the Gibbs paradox: both the thermodynamic entropy of mixing and the Gibbs free energy change are zero during the formation of any ideal mixtures. Information loss is the driving force of these spontaneous processes. Information is defined as the amount of the compressed data. Information losses due to dynamic motion and static symmetric structure formation are defined as two kinds of entropies – dynamic entropy and static entropy, respectively. There are three laws of information theory, where the first and the second laws are analogs of the two thermodynamic laws. However, the third law of information theory is different: for a solid structure of perfect symmetry (e.g., a perfect crystal), the entropy (static entropy for solid state) $S$ is the maximum. More generally, a similarity principle is set up: if all the other conditions remain constant, the higher the similarity among the components is, the higher the value of entropy of the mixture (for fluid phases) or the assemblage (for a static structure or a system of condensed phases) or any other structure (such as quantum states in quantum mechanics) will be, the more stable the mixture or the assemblage will be, and the more spontaneous the process leading to such a mixture or an assemblage or a chemical bond will be.

**Keywords:** Information Theory, Entropy of Mixing, Symmetry, Stability, Indistinguishability.

**PACS:** 65.40.gd; 89.70.Cf; 87.19.lo; 05.70.Ln; 47.20.Hw

## 1. INTRODUCTION

A paper entitled "The Gibbs paradox" was presented by Jaynes at one of the MaxEnt conferences [1]. Gibbs' paradox has been also a topic of discussion recently (see e.g., [2] or http://www.mdpi.org/lin/entropy/gibbs-paradox.htm for a collection of papers). As a chemist, the present author has been never satisfied by any of these solutions or explanations (see a critical review prepared and published in Chinese almost 20 years ago [3]). A very simple experimental resolution of Gibbs paradox will be given by the author, following the observation of thermodynamic and physicochemical measurements. A more general consideration within information theory leads to a higher entropy-higher similarity relationship. A new kind of entropy is defined and called static entropy.

Jaynes, after reviewing the history, said that the conceptual difficulties are the main factor to greatly hinder the further development of the science [4]; this is also exactly the same difficult situation surrounding the resolution of Gibbs paradox. It has been one of the main tasks of the present author to investigate the concepts of similarity, indistinguishability, symmetry, order, information loss and stability and to clarify their relationships.

## 2. GIBBS PARADOX

For the mixing of two substances A and B, the Gibbs paradox states that if the two substances are identical, there will be no entropy change, yet the slightest difference between the two will yield a considerable entropy change, the entropy of mixing [1,2] for mixing 1 mol of A and 1 mol of B,

$$\Delta S_T = 2R \ln 2 = 11.53 \text{ JK}^{-1} \quad (1)$$

In other words, the entropy of mixing is not a continuous function of the degree of difference between the two substances in the Gibbs paradox.

As illustrated in Figure 1, two gases (for example, two chiral 2-deuteroethanol molecules A and B, as examples of two low pressure gases which can be treated as ideal gases) are mixed and there will be entropy of mixing (Eq. 1) and Gibbs free energy of mixing, even though these two molecules are of the same mass, have very similar properties and their difference cannot be recognized by any heat engine.

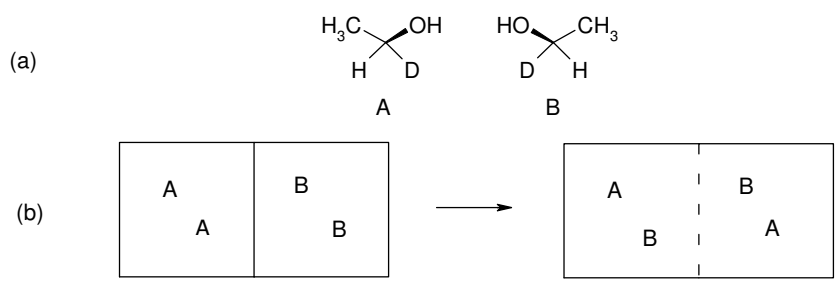

**FIGURE 1.** Mixing deuterated ethanols as two chiral molecules A and B.

John von Neumann [5] provided a resolution of Gibbs paradox. Suppose two kind of particles are represented by the states $|\phi_A\rangle$ and $|\phi_B\rangle$, respectively, and the overlap integral $\langle \phi_A | \phi_B \rangle = a$ are calculated. Von Neumann's entropy of mixing formula is

$$\begin{aligned} S_{mix} &= 2\log_2 2 - (1+a)\log_2(1+a) - (1-a)\log_2(1-a) \\ &= 2 - (1+a)\log_2(1+a) - (1-a)\log_2(1-a) \end{aligned} \quad (2)$$

which is depicted in Figure 2. The maximum value of entropy of mixing is 2 bits.

Thermodynamic entropy of mixing (Eq. 1) and Gibbs free energy of mixing have been calculated according to the Gibbs Paradox statement [6]. A calorimeter or any other equipment might be employed to determine the change of thermodynamic functions $\Delta S_T$ or $\Delta G = -T \Delta S_T$ (where $G$ is the Gibbs free energy, $T$ is temperature) of the mixing process (Eq. 1). Unfortunately energy changes have never been observed for the formation of ideal mixtures. This leads to the conclusion that the *thermodynamic entropy change $\Delta S_T$ of a typical isothermal, isobaric process of ideal mixture formation is always zero*, whether the components are different or identical. This conclusion can be taken as an experimental resolution of Gibbs paradox [7]. It is clear that entropy of mixing has nothing to do with energy. A mixing process is a process of information loss which can be pertinently discussed only in the realm of information theory and entropy of mixing is an (information theory) entropy. Chemical sensors or biosensors can be used to assess the information loss during the mixing process. Mixing 1 mol of A and 1 mol of a different substance B will results in an increase of at most 2 bits of (information theory) entropy if the two parts of the gas container are used to record 2 bits of data (Figures 2 and 3). If the entropy is regarded as an (information theory) entropy and the microscopic structure details of the involved condensed phases, including the containers, are ignored (*vide infra*), von Neumann's resolution as illustrated in Figure 2 appears valid [5] because there is no information to be lost if A and B are the same and less information to be lost if A and B are similar. If A and B are different, there are full amount of information to be lost, see Figure 3 (The calculation details can be found in the next two sections).

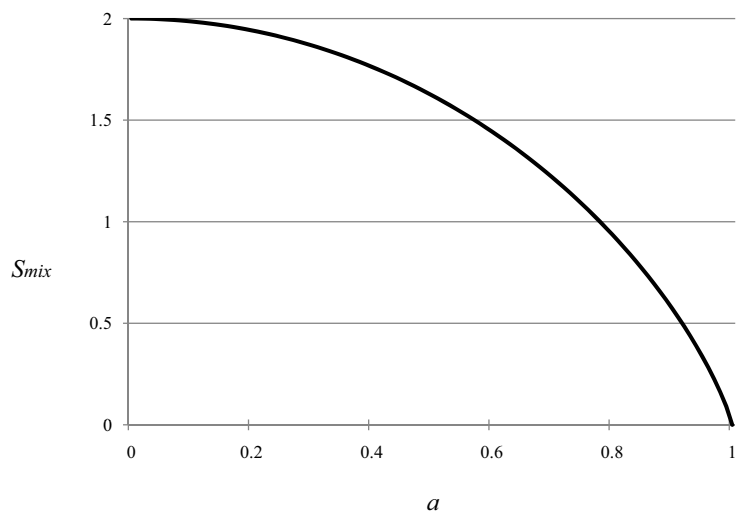

**FIGURE 2.** Entropy of mixing ranges continuously from 2 bits (orthogonal states, $a=0$) to 0 (identical states, $a=1$).

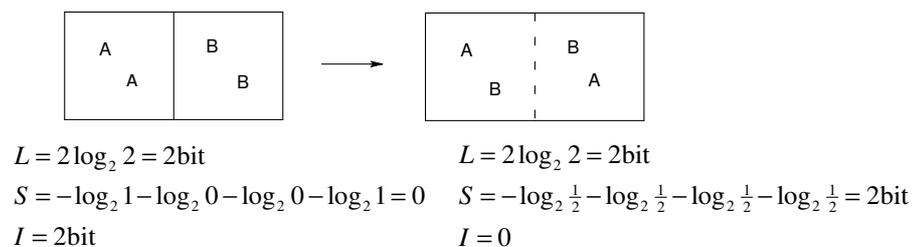

$L = 2\log_2 2 = 2\text{bit}$         $L = 2\log_2 2 = 2\text{bit}$

$S = -\log_2 1 - \log_2 0 - \log_2 0 - \log_2 1 = 0$   $S = -\log_2 \tfrac{1}{2} - \log_2 \tfrac{1}{2} - \log_2 \tfrac{1}{2} - \log_2 \tfrac{1}{2} = 2\text{bit}$

$I = 2\text{bit}$           $I = 0$

**FIGURE 3.** Total amount of data, entropy and information before and after the mixing of a system of two different ideal gases in two parts of a container, $M=2$, $N=2$ (*vide infra*).

Generally, however, instead of the word "mixing", the word "merging" can be used for the process of combining several parts of substance originally in several containers and the containers (E.g., the water in the tube, Figure 4a) themselves are also considered, for heterogeneous systems (Figure 4). For the solid mixture (Figure 4b), A (labeled as 0) is the container of B (labeled as 1) and *vice versa*. Then, it is always a merging process, whether the substances are very different or very similar or even the same.

The conventional way of entropy of mixing calculation would predict that the mixing (or merging) process of different (distinguishable) substances is more spontaneous than the merging process of the same (indistinguishable) substances. However, this contradicts all the observed facts in the physical world where the merging process of the identical substances is the most spontaneous one; immediate examples are spontaneous merging of oil droplets in water (Figure 4a) and spontaneous crystallization where the indistinguishable joined lattice cells assemble together (Figure 4b) [7-12]. More similar substances are more spontaneously miscible. The two alcohols methanol and ethanol are miscible because they are very similar.

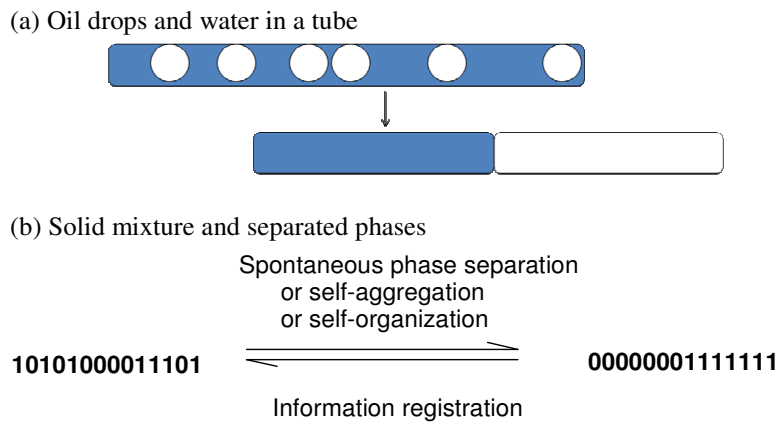

**FIGURE 4.** Two cases where the information loss is observed during the merging of the phases which results in phase separation. On the left side of Figure (b) the molecules A and B labeled as 1 and 0 are mixed and appear as a one-dimensional solid mixture or a binary string.

As shown in Figure 4b, starting from a binary solid mixture, the process of merging 1 mol of molecules A to become one phase and merging of 1 mol of molecules B to form another phase leads to a tremendous information loss or (information theory) entropy increase of

$$\Delta S = 2 \times 6.022 \cdot 10^{23} \, \text{bit} = 1.506 \cdot 10^{23} \, \text{byte} \tag{3}$$

where $6.022 \cdot 10^{23}$ is Avogadro's number; and there will be at most only 2 bits of information ($I$) left (see Figure 4b). This simple way of information loss assessment will be explained in the following sections.

## 3. INFORMATION DEFINED AS COMPRESSED DATA

There are many definitions of information. A new definition of information is given as the compressed data [7]. For a definition of data compression, see http://en.wikipedia.org/wiki/Data_compression. The compressed data $I$ has less number of bits than the total amount of data $L$:

$$0 \leq I \leq L \tag{4}$$

There is a disadvantage in this definition because the noise recorded as data cannot be compressed (this will be discussed in the following section). It also depends on the compression software to be used. Despite of these drawbacks, this definition is the most plausible one because the data can be compressed via a standard algorithm and this gives an objective, reliable quantitative relation of information ($I$), the amount of data ($L$) and entropy ($S$).

Lewis' entropy-information relation (gain of entropy means loss of information) [13,14] is adopted:

$$\Delta S = -\Delta I \tag{5}$$

The total amount of data $L$ is the sum of entropy and information,

$$L = S + I \tag{6}$$

Great attention should be paid to the fact that the other very popular relation of entropy and information "entropy is a measure of information" (see ref. 15, also, p.37, ref. 4) is not accepted. Nevertheless, Lloyd, a physicist, pointed out that "conventional information theory defines the total amount of information registered by a system to be the difference between the system's actual entropy and its maximum possible entropy" [16]:

$$I = L - S \tag{7}$$

where the total amount of data $L$ is also the maximum possible entropy (or the maximum possible information) [7,12].

Following this definition and the Lewis relation of entropy and information, three laws of information theory can be proposed [12], with the first two given here first:

> *The first law of information theory:* the total amount of data $L$ (the sum of entropy and information, $L = S + I$) of an isolated system remains unchanged.
>
> *The second law of information theory:* Information $I$ of an isolated system decreases to a minimum at equilibrium.

If entropy change is exactly the information loss (Eq. 5), the conservation of $L$ can be very easily satisfied,

$$\Delta L = \Delta S + \Delta I = 0. \tag{8}$$

The third law will be given in the following section.

The information theory concepts can be introduced to thermodynamics and statistical mechanics. Thermodynamic entropy $S_T$ will be easily represented in the more general (information theory) entropy $S$

$$S_T = kS \tag{9}$$

where $k$ is the Boltzmann constant. The total amount of data $L$ and Information $I$ may also be easily introduced to thermodynamics and statistical mechanics:

$$E = kTL \tag{10}$$
$$G = kTI \tag{11}$$

where $E$ is the total energy. As Schrödinger remarked, he would use the word free energy $G$ instead of negative entropy (cited on p.254, ref. 6).

A chemical system of heterogenerous substances in the solid state can be taken as a simple discrete case [15] in information theory. The number of substances is $M$. These $M$ kinds of substances (chemical species) are distributed at $N$ sites. When we discuss the written language, $M$ becomes the number of different characters. A text is the distribution of these characters written on $N$ sites by typesetting and the total amount of data is

$$L = N \ln M. \tag{12}$$

For such system, the total number of microstates (or the maximum number of different texts) will be

$$w = M^N. \tag{13}$$

The binary string given in Figure 4b has $M=2$, $N=14$. Binary systems will be often taken as the model system in this paper and the total amount of data is

$$L = \ln 2^N \text{ (nat)}$$
$$= \log_2 2^N = N \text{(bit)} \tag{14}$$

An elegant description of such a binary system can be found in Chapter 1 of the excellent book of Kittel and Kroemer [17].

Incidentally, scientific research may be regarded as a series of data compressions during the three stages of *data–information–knowledge* progress, either manually or assisted by computer. For example, the largest prime number at this moment given at http://www.mersenne.org/prime10.txt can be represented in different number systems and character sets, see Table 1. It is interesting to note that if a book of 300 pages in English (character set has $M=256=2^8$) were translated into Chinese (the minimum character set $M=4096=2^{12}$), the published Chinese version would be only 200 pages because $2^{12x}=2^{8\times300}$ and $x=200$. The meaning of diversity has been discussed in detail [9]. The mentioned prime number should have 4,072,832 bytes if this number is represented using all the available 256 characters available for a text file. If the data have been stored in a hard disk with $L=100$ MB, this will stay unchanged and the information is $I=4,072,832$ bytes. Finally it can be expressed as a formula with the amount of data only a few bytes and shorter than one line:

$$2^{32,582,657} - 1 \tag{15}$$

as a Mersenne Prime (see: http://en.wikipedia.org/wiki/Largest_known_prime). Of course this final formula representing this number has the data tremendously compressed. Maybe many large primes can be expressed as a concise formula and this might be a task for the mathematician. In the next section, we will show that symmetries lead to data compression.

**TABLE 1.** The Data of the Largest Prime Number. This Number Can Be Represented by Using Different Character Set and Recorded as a Text File.

| Character set | M | N | Number of pages |
|---|---|---|---|
| binary | 2 | 32582660 | 3258 |
| decimal | 10 | 9808358 | 980 |
| hexadecimal | 16 | 8145665 | 814 |
| ISO/IEC 8859-1 | 256 | 4072832 | 407 |

Among the $2^N$ microstates (Figure 5), following the Gaussian distribution, the number of the most probable distributions with equal number of 0 and 1 is:

$$\frac{N!}{(0.5N)!(0.5N)!} \tag{16}$$

Some of them are noise-like texts (or microstates) and can hardly be compressed within the binary system (They can be represented by using other character set and the data can be further reduced, as shown in Table 1. We do not discuss this further in detail). Some of these $2^N$ microstates (or texts), however, can be easily compressed due to symmetry, see for example Figure 5c which has very high data compressibility even though there are equal number of 0 and 1. As shown in Figure 5b, "10" is repeated and the data should be highly compressible. It would be an interesting research task to find the compressibility distribution of these $2^N$ microstates. Kolmogorov complexity or algorithmic complexity studies might be related to this topic [18]. The structure with the pure phases (Figures 5d and 5e), of course, can be compressed by inspection because they possess obviously the highest possible symmetries.

(a) 010011010110101000110

(b) 101010101010101010101010

(c) 00000000001111111111

(d) 00000000000000000000

(e) 11111111111111111111

**FIGURE 5.** Illustration of the symmetry and the data compression.

## 4. SYMMETRY, INFORMATION LOSS AND DATA COMPRESSIBILITY

Several examples are given (Figure 6) to illustrate that dynamic motions can increase the symmetry and lead to information loss [8]. Pauling's residual entropy of ice is another pertinent example: residual entropy can be calculated by considering the increased symmetry due to the hydrogen bond [19]. In Pauling's argument, the higher local symmetry ($T_d$) in ice due to hydrogen-bonding (a dynamic motion of H atom between two O atoms) provides a positive contribution to the value of the residual entropy.

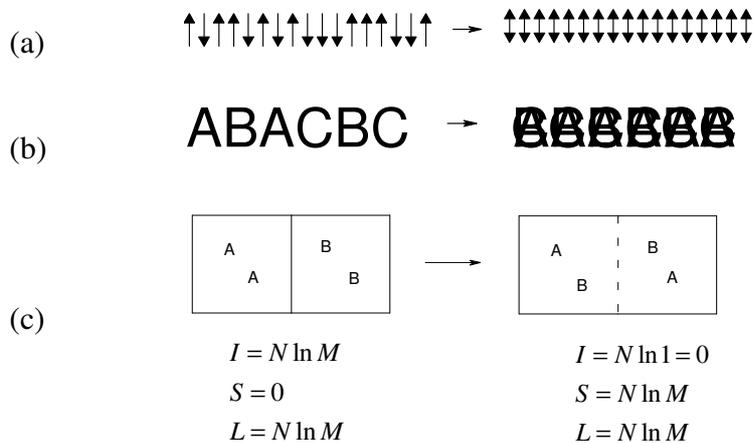

**FIGURE 6.** Dynamic symmetries and information loss. The left side structures are nonsymmetric. The right side structures are symmetric [12]. For figure (b), $M=3$.

Two examples of microstates with different static symmetries and the calculation of static entropies [7,8,12] are given in Figure 7. Definitely there is a fundamental flaw in the existing theories of the famous Ising model for the ferromagnetic phenomena because it should be unlikely to form the frequently observed unique all-spin-up structure (right side of Figure 7a) as the probability is extremely small: $1/(2^{10^{23}})$.

Another problem is the spin-spin interaction energy: one of the most probable microstates should be the one shown on the left side of Figure 7a which has the lower energy, while the spin-spin interaction energy is the highest at the all-spin-up structure (see the right side of Figure 7a). Now we have static entropy [7,8,12] as information loss due to symmetry and can act as the driving force for the formation of the ferromagnetic structure.

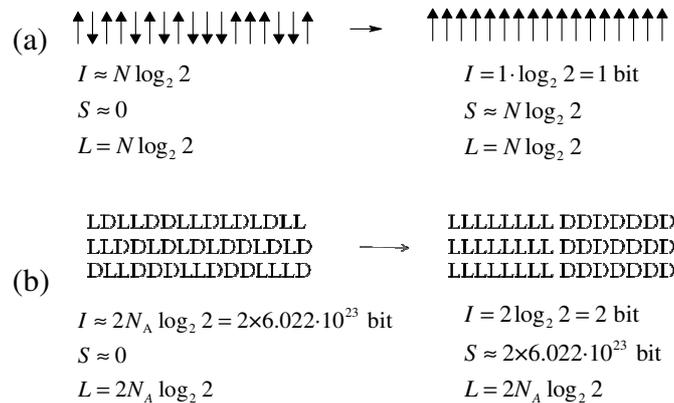

**FIGURE 7.** Static symmetries and information loss. (a) Magnetic spins. The right side is symmetric ferromagnetic structure. (b) Separation of two kinds of chiral molecules.

Why is the influence of the static entropy or the information loss due to the symmetry is so strong that it can overcome the disadvantage of the increased energy at the high-spin structures? This remains an open question.

Symmetry leads to high data compressibility during the data analysis. For example, a large number of noisy images or noisy spectra can be collected and compressed through superimposition, where the weak signals appear *repeatedly*. The observation and storage of the data of a sinusoidal signal in the time domain can occupy as much hard disk space as one can provide. The periodicity – a symmetry – provides the possibility to carry out a Fourier transform to have a much reduced amount of data in the frequency domain. Halley's Comet is an astral body that can be seen and recorded every 75.3 years. The data with the repetition symmetry were "compressed" by Edmond Halley as a phenomenon observable every 75.3 years.

The relation of symmetry-information loss (data compressibility) for static or solid structures presented here can be summarized as the third law of information theory [12]:

> *The third law of information theory:* for a perfect symmetric static structure, the information $I$ approaches zero and the static entropy $S$ is the maximum.

This law conforms very well to the Curie-Rosen symmetry principle [12,20,21].

Any static symmetry will define a static entropy as one kind of (information theory) entropy which is related to stability. The third law of thermodynamics is still valid if the considered entropy is the thermodynamic entropy $kS$ or dynamic entropy defined in this paper. The discussions about the symmetry numbers in special cases and the correct explanation of experimental findings, such as the estimation of the meting points [22,23], will be discussed elsewhere.

# 5. SIMILARITY PRINCIPLE

Symmetry is the highest value of similarity [8]. The symmetry–information loss relation can be further generalized as follows:

> *The Similarity Principle*: If all the other conditions remain constant, the higher the similarity among the components is, the higher value of entropy of the mixture (for fluid phases) or the assemblage (for a static structure or a system of condensed phases) or any other structure (such as chemical bond or quantum states in quantum mechanics) will be, the more stable the mixture or the assemblage will be, and the more spontaneous the process leading to such a mixture or an assemblage or a chemical bond will be.

Based on the second law and the third law of information theory and the fact that symmetry is the highest value of similarity, the Curie-Rosen symmetry principle [20] can be regarded as a special case of similarity principle. The similarity principle can be proved by using Gibbs' inequality [8,12]. The condition for the maximum entropy must be the property indistinguishability among the $w$ microstates so that the probabilities are of the same value: $1/w$. As commented by Jaynes, a perfectly symmetric die ought to show both faces equally often [page 62 of ref.4].

Several kinds of indexes have been developed by mathematical chemists. One of them, the chiral index, has been reported by Michel Petitjean at the MaxEnt2008 conference. These kinds of indexes can be used to define similarity (for example, see ref.24). As another example, the similarity of molecular masses $m_A$ and $m_B$ is considered [25].

Entropy, similarity and stability consideration can be applied to a molecular system involving different energy states [26]. For the formation of chemical bond, as Pauling described, "…the two structures with the same energy in resonance make equal contributions to the normal state of the system" and form the strongest covalent bond [27]. The familiar Boltzmann factor [26] can be taken as a similarity scale, see Figure 8. In this contest, it can be mentioned that life is manifested at the ambient temperature of 300K or about 25°C and avoids both high dynamics entropy (high $T$) and high static entropy (low $T$) as illustrated in Figures 6 and 7, respectively. A living organism is a heterogeneous system with driving force of both dynamics entropy trap and static entropy trap.

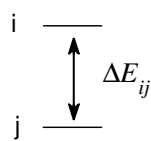

Boltzmann factor $f_{ij} \propto e^{-\frac{\Delta E_{ij}}{kT}}$

$\rightarrow 1$ at high $T$, dynamic entropy (dynamic symmetry)

$\rightarrow 0$ at low $T$, static entropy (static symmetry)

**FIGURE 8.** System with two states with a fixed energy level difference.

Entropy is a monotonically increasing function of similarity [7-12] which may be changed by increasing temperature (Figure 8). Let us consider a practical example of miscibility experiment which is very pertinent because we are discussing theory of mixtures. Suppose there are two liquids A (water, $H_2O$, for example) and B (an alcohol ROH with a bulky group R) originally separated by solid wall. The miscibility studies can be carried out by replacing the wall with a porous wall and the two liquids are allowed to mix (see Figure 9). The two parts are used as two positions for the binary system A,B to record data, $L = \log_2 2^2 = 2$ bit. Suppose the volumes do not change when they mix at any volume ratio and the volume concentrations are in the range of 0 to 1. Then, set $c_{BL} = c$ as the found volume concentration of B molecules that transferred to the left side which is originally occupied by pure A before the solid wall is replaced. The same volume of A will enter the right side of the porous wall and $c_{AR} = c, c_{BR} = 1 - c$ as the found volume concentrations of A and B, respectively. Temperature $T$ is changed to tune the similarity between A and B and the miscibility is measured by chromatographic method. At certain low temperature, they remain separated: $c_{AL} = 1$ and $c_{BL} = 0$ on the left side of and $c_{AR} = 0$ and $c_{BR} = 1$ on the right side. This gives $S = 0$. If half of B entered the left side from the right side, $c_{AL} = c_{BL} = c_{AR} = c_{BR} = 0.5$, indicating the highest value of miscibility and correspondingly the highest value of entropy: $S = 2$. A similarity (also the miscibility) between A and B can be defined as $z = 2c$. Entropy of mixing is calculated using Eq. 17 and is plotted in Figure 10.

$$\begin{aligned} S &= -c_{AL} \log_2 c_{AL} - c_{BL} \log_2 c_{BL} - c_{AR} \log_2 c_{AR} - c_{BR} \log_2 c_{BR} \\ &= 2\left[-c \log_2 c - (1-c) \log_2 (1-c)\right] \\ &= 2\left[-0.5z \log_2 (0.5z) - (1-0.5z) \log_2 (1-0.5z)\right] \end{aligned} \quad (17)$$

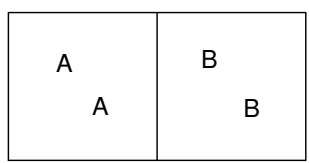
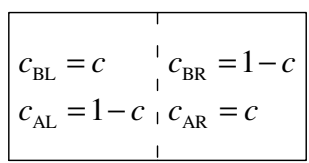

(a) Before the mixing experiment.   (b) The two parts are allowed to mix.

**FIGURE 9.** The miscibility experiment is done by replacing the solid wall in Figure (a) with a porous wall followed by the concentration measurement in the two parts.

For the miscibility experiment concerned, the similarity between A and B are the highest if the volume ratio in the ideal mixture is 1:1. The two chiral 2-deuteroethanols A and B (Figure 1a) can become indistinguishable in many cases: for a heat engine and for the observer who are doing miscibility experiment. Of course all the ideal gases are indistinguishable if the miscibility is considered. Ideal gases mix not because they are different, but because as ideal gases they are very similar. Two phases (figure 4a, for example) separate because the very similar substances would like to merge. All these processes, whether at higher temperature or at lower temperature, are governed by the rule that the higher the similarity among the components is, the higher the value of entropy (static or dynamic or both) will be and the higher the stability of the final structure will be.

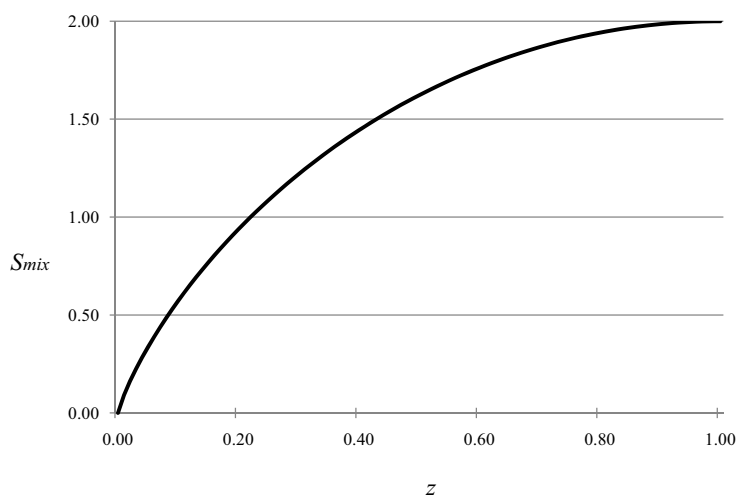

**FIGURE 10.** Entropy of mixing and miscibility.

Finally the second law of information theory can be given yet another very general form:

> *The second law of information theory:* The similarity of the components or parts will increase to the maximum, or *the difference cannot increase spontaneously*.

## 6. CONCLUSIONS

As a more general resolution of Gibbs paradox, entropy is found to be a monotonically increasing function of the similarity. A new entropy-similarity relationship called similarity principle has been set up. We also revised the information theory where the three laws of information theory are given. Similarity principle is more general than the Curie-Rosen symmetry principle. Information loss alone can be the driving force of a physicochemical process where energy minimization rule does not play any role. It is certain that these new and clear relations of similarity, symmetry or indistinguishability, information loss, entropy and stability can be applied for reformulating the theoretical foundation of theoretical physics, physical chemistry, and biophysics.

## ACKNOWLEDGMENTS

Prof. Dr. Carlos Alberto de Bragança Pereira kindly provided the opportunity for the author to present a lecture and to write this chapter. The author is also very grateful to his long time colleague Dr. Derek McPhee for his collaboration and assistance. Derek corrected the English for this paper. Dietrich Rordorf prepared figures 2 and 10.